\documentclass[conference]{IEEEtran}
\IEEEoverridecommandlockouts

\usepackage{amsmath,amssymb,amsfonts}
\usepackage{algorithmic}
\usepackage{graphicx}
\usepackage{microtype}
\usepackage{subcaption}
\usepackage{caption}
\usepackage{textcomp}
\usepackage{xcolor}
\usepackage{cite}
\usepackage{tabularx}
\usepackage{booktabs}
\usepackage{tikz}
\usepackage{url}
\usepackage[prependcaption,textsize=scriptsize]{todonotes}
\definecolor{mycolor}{HTML}{FF6600}
\definecolor{coolcolor}{HTML}{26209e}
\definecolor{plancolor}{HTML}{aaa6ed}
\definecolor{nelcolor}{HTML}{3060F2}

\usetikzlibrary{shapes.geometric, arrows, calc, fit, shapes}
\tikzstyle{empty} = []

\newcolumntype{C}{>{\centering\arraybackslash}X}
\newcolumntype{L}{>{\raggedright\arraybackslash}X}
\newcolumntype{R}{>{\raggedleft\arraybackslash}X}
\newcolumntype{P}[1]{>{\raggedright\arraybackslash}p{#1}}

\newcommand{\tablesep}{\vspace{1pt}}

\begin{document}

\title{Recovering the Zipfian Distribution in \\ Unsupervised 
Term Discovery\\
\thanks{
This work is supported by the Het Jan Marais Fonds.
}
}

\author{
\IEEEauthorblockN{Danel Slabbert, Simon Malan, Herman Kamper}
\IEEEauthorblockA{
\textit{Electrical and Electronic Engineering}, \textit{Stellenbosch University}, South Africa \\
24051055@sun.ac.za, 24227013@sun.ac.za, kamperh@sun.ac.za
}
}

\maketitle 

\begin{abstract}
Unsupervised term discovery involves segmenting unlabelled speech into word- or syllable-like units and clustering these into a lexicon of candidate types.
True lexicons follow a Zipfian distribution, yet the dominant centre-based clustering approach---K-means---produces a more uniform distribution due to an inductive bias toward spherical clusters.
In this paper we revisit graph-based clustering as a bottom-up alternative, where segment embeddings are connected by pairwise similarity and partitioned using the Leiden algorithm.
We show that graph clustering substantially outperforms centre-based approaches (K-means, GMM, BIRCH) in both word- and syllable-level lexicon discovery across three languages, producing more Zipf-like distributions. 
Another bottom-up approach, agglomerative clustering with average linkage, also performs well, although it is computationally less efficient and allows for less control over the resulting distribution.
Our work calls into question the dominance of centre-based clustering for term discovery, and promotes graph clustering as an attractive alternative.
\end{abstract}


\section{Introduction}
Unsupervised term discovery builds a lexicon from unlabelled speech by identifying boundaries and then clustering the resulting segments into phonetically homogeneous groups~\cite{Ludusan2014}.
Although human infants achieve this within their first year of life~\cite{Saffran1996, Jusczyk2002, Bergelson2012}, it remains a challenging problem for computational systems.
The difficulty stems from the absence of explicit word delimiters and the acoustic variability across instances of the same word~\cite{Rsnen2012}.
Developing models that approximate this ability can inform theories of language acquisition~\cite{Dupoux2018} and support speech technologies in low-resource settings~\cite{Besacier2014}.

Early approaches for term discovery identified recurring acoustic patterns using dynamic time warping and clustered them via a similarity graph~\cite{Park2008, Lyzinski2015}.
However, they relied on conventional acoustic features that capture surface properties of the signal.
Self-supervised learning (SSL) changed this: models trained on large quantities of unlabelled speech learn representations that are far more robust to speaker and acoustic variability while encoding richer phonetic structure~\cite{Hsu2021, Chen2022}. 
Modern discovery systems therefore segment and cluster speech in learned embedding spaces~\cite{Lee2015, Rsnen2015, Kamper2017es, Bhati2020, Bhati2021, Algayres2022, Cuervo2022, Okuda2023, Kamper2023, vanNiekerk2024}.
\begin{figure}[b!]
    \captionsetup{font=small}      
    \centering
    \vspace{-10pt}  
    \includegraphics[width=\linewidth]{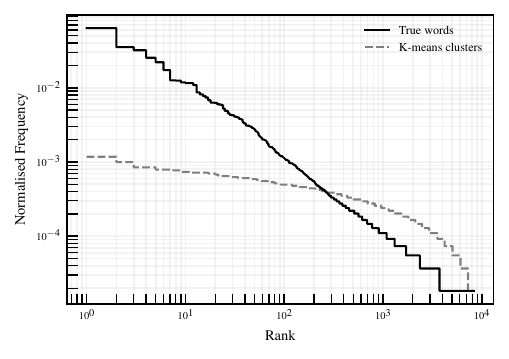}
    \vspace{-20pt}
    \caption{
        True word-type distribution vs K-means cluster frequencies over true word segments using averaged WavLM features on LibriSpeech dev-clean.
        K-means produces a near-uniform distribution, failing to capture the long-tailed Zipfian structure of natural language. 
    }
    \label{fig:zipf_plot_words_dev_clean}
\end{figure}

Utilising SSL features, most discovery systems now first segment speech before clustering the resulting segments.
For clustering, K-means has seen widespread adoption due to its simplicity and efficiency~\cite{Pasad2024,Malan2024}.
However, K-means has an inductive bias toward similarly sized clusters.
As illustrated in Fig.~\ref{fig:zipf_plot_words_dev_clean}, the resulting K-means distribution is therefore poorly aligned with the long-tailed Zipfian type-frequency distributions characteristic of natural language~\cite{Piantadosi2014}.
Other common clustering approaches~\cite{Zhang1996, Blimes1998, Bishop2006} also favour compact clusters around representative centres.

This paper investigates whether we can recover the Zipfian distribution in the lexicons of unsupervised term discovery systems by using a clustering method with a different inductive bias.
To this end, we compare three centre-based clustering methods (K-means, BIRCH, GMMs) with two bottom-up alternatives (graph clustering, agglomerative clustering).
Graph clustering---which was used in the earliest term discovery work~\cite{Park2008}---is particularly attractive because its hyperparameters provide control over the properties of the resulting lexicon.
In addition, we evaluate the five clustering methods under different segmentation granularities to assess how this affects lexicon induction: ground-truth word boundaries, providing an upper bound; ground-truth syllable boundaries, which evaluate clustering over shorter, more uniform units; and an unsupervised syllabic segmentation, reflecting a fully unsupervised setting with noisy segments.
We are particularly interested in the syllabic units because these can generally be segmented more reliably than words by unsupervised systems\cite{Rsnen2015}, and because they have seen growing adoption in spoken language modelling~\cite{Cho2025, Baade2025, Visser2026}.

In all of the above, we use SSL embeddings trained on English data and perform experiments on English.
To match a real zero-resource setting, we repeat our comparisons on Afrikaans and French data without altering the system settings. 

Our contributions are threefold:
(1)~We show that the choice of clustering method plays a crucial role in unsupervised lexicon induction: bottom-up methods consistently outperform centre-based clustering across segmentation conditions and evaluation languages.
(2)~We analyse the distributional properties of induced lexicons and show that, compared to centre-based approaches like K-means, bottom-up methods better recover the long-tailed Zipfian distribution of word-level lexicons while also adapting to the less skewed distributions observed for syllabic units.
(3)~We show that graph clustering is a particularly useful bottom-up approach with two interpretable 
hyperparameters---a similarity threshold and a resolution parameter---that provide control over the induced type-frequency distribution.

Code is provided at:~\url{https://github.com/adendorffy/zipf-clus}.

\begin{figure}[!t]
    \captionsetup{font=small}      
    \begin{tikzpicture}[node distance=2cm]
    \node (n1) [empty, anchor=west] at (0,0) {};
    \node[anchor=south, font=\small] at ($(n1.north) + (-1.4, 1.3)$) {(a) Segmentation};
    \begin{scope}[shift={([xshift=-2.5cm, yshift=-0.5cm]n1.center)}, xscale=0.3, yscale=0.4]

        \draw[thick]
        (0,0) -- (0.1,0.05) -- (0.2, -0.1) -- (0.3, 0.2) -- (0.4, -0.4)
        -- (0.5, 0.7) -- (0.6, -0.9) -- (0.7, 1.0) -- (0.8, -1.5) -- (0.9, 0.7)
        -- (1.0, -0.7) -- (1.1, 0.5) -- (1.2, -0.4) -- (1.3, 1.2) -- (1.4, -1.8)
        -- (1.5, 2.1) -- (1.6, -1.9) -- (1.7, 1.5)  -- (1.8, -0.8) -- (2.0, 0.2)
        -- (2.1, -0.05) -- (2.2, 0) 
        -- (2.3, 0.02) 
        -- (2.4, 0.6) -- (2.5, -1.1) -- (2.6, 2.1) -- (2.7, -1.9) -- (2.8, 2.3) 
        -- (2.9, -2.0) -- (3.0, 1.4) -- (3.2, -1.8) -- (3.4, 2.4) -- (3.6, -2.5)
        -- (3.8, 2.1) -- (4.0, -1.5) -- (4.2, 0.8) -- (4.4, -0.3) -- (4.7, 0.1)
        -- (4.9, 0)
        -- (5.0, -0.01) 
        -- (5.1, 0.5) -- (5.3, -1.2) -- (5.5, 1.8) -- (5.7, -2.0) -- (5.9, 2.2)
        -- (6.1, -1.7) -- (6.3, 1.1) -- (6.5, -0.9) -- (6.7, 1.4) -- (6.9, -1.6)
        -- (7.1, 1.2) -- (7.3, -0.4) -- (7.5, 0.2) -- (7.7, -0.05) -- (7.9, 0)
        -- (8.0, 0.02);

    \end{scope}
    \draw[dashed, thin] (-2.4, -1.5) -- (-2.4, 0.5);
    \draw[dashed, thin] (-1.75, -1.5) -- (-1.75, 0.5); 
    \draw[dashed, thin] (-1.0, -1.5) -- (-1.0, 0.5);
    \draw[dashed, thin] (0.0, -1.5) -- (0.0, 0.5);

    \node[anchor=south, font=\footnotesize] at (-1.1, -2.4) {Boundaries};

    \begin{scope}[yshift=0.5cm] 
    
        \draw[fill=white] (-2.4, 0.4) rectangle (-1.85, 0.8);
        \draw[fill=white] (-1.75, 0.4) rectangle (-1.1, 0.8);
        \draw[fill=white] (-1, 0.4) rectangle (-0.1, 0.8);

    \end{scope}

    \node (n2) [empty, right of=n1,  minimum width=4cm, node distance=1cm] {};
    \node[anchor=south, font=\small] at ($(n2.north) + (0.7, 1.3)$) {(b) Representation};
    \draw[thick, dashed, ->] ($(n1.west) + (0.1, 1.1)$) -| ($(n1.west)!0.5!($(n2.east) + (-2.0, 0.5)$)$) |- ($(n2.east) + (-2.1, -0.2)$);
    \node[anchor=south, font=\footnotesize] at ($(n2.north) + (0.8, 0.7)$) {(Per segment)};

    \foreach \i in {1,...,3} {
        \coordinate (rect\i) at ($(n2.north west) + (2.0 + \i*0.4, -0.4)$);
        \draw[fill=white] ($(rect\i) + (-0.09,-0.3)$) rectangle ++(0.3, 1.2);
    }

    \foreach \i in {1,...,3} {
        \draw[thin, ->] ($(n2.north west) + (2.05 + \i*0.4, -0.85)$) -- ++(0, -0.3);
    }

    \foreach \i in {1,...,3} {
        \draw[fill=white] ($(n2.north west) + (2.05 + \i*0.4, -1.4)$) circle (0.1cm);
    }
    \draw[thin, ->] ($(n2.north west) + (2.85, -1.7)$) -- ++(0, -0.3);
    \draw[fill=gray] ($(n2.north west)+(2.85, -2.3)$) circle (0.1cm);
    \node[anchor=south, font=\footnotesize] at ($(n2.north west)+(2.4, -2.55)$) {$y_{i,j}$};

    \node (n3) [empty, right of=n2, minimum width=4cm, node distance=2cm] {};
    \node[anchor=south, font=\small] at ($(n3.north) + (1.5, 1.3)$) {(c) Clustering};  

    \begin{scope}[shift={($(n3.north) + (2.0, -2)$)}]
        \draw[<->, thick] (-1.3, 2.5) -- (-1.3, -0.2) -- (1.0, -0.2);

        \foreach \p in {(-0.1, 1.4), (-0.6, 1.4), (-0.8, 2.0), (-0.3, 2.2)} {
            \fill[gray] \p circle (0.1);
        }
        \draw[dashed, gray!40] (-0.5, 1.8) circle (0.7); 
        \node[font=\tiny] at (-0.3, 2.7) {Cluster 1};

        \foreach \p in {(0.0, 0.15), (0.0, 0.8), (0.5, 0.8), (0.5, 0.3)} {
            \fill[gray] \p circle (0.1);
        }
        \draw[dashed, gray] (0.2, 0.5) circle (0.6); 
        \node[font=\tiny] at (-0.7, 0.1) {Cluster 2};
        
        \fill[black] (-0.5, 1.8) circle (0.11);
        \fill[black] (0.2, 0.5) circle (0.11);
    \end{scope}
    \draw[thick, dashed, ->] ($(n2.north west)+(3.2, -2.3)$) |- ($(n2.north west)+(4.0, -2.3)$) |- ($(n2.north west)+(4.0, -1.1)$) -- ($(n3.east)+(-1.5, -1.0)$);
    
    \end{tikzpicture}
    \caption{The typical unsupervised term discovery pipeline: 
    (a)~speech audio is partitioned into segments; 
    (b)~each segment is mapped to a representation using an SSL model; 
    and (c)~a clustering algorithm groups segments into a lexicon of types.
    }
    \vspace{-10pt}
    \label{fig:pipeline}
\end{figure}
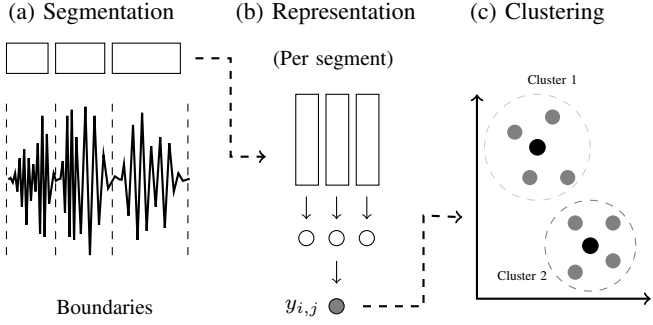

\section{Methodology}
We follow the three typical stages of unsupervised term discovery, as illustrated in Fig.~\ref{fig:pipeline}, keeping the representation stage fixed across all experiments while varying the segmentation levels and clustering methods.
This allows us to assess how different clustering assumptions and segmentation granularities affect lexicon quality and structure.

\subsection{Segmentation and representation}
Given a segmentation of the input speech by some unsupervised method (Fig.~\ref{fig:pipeline}a), we define each segment $x_{i:j}$ by its start and end frames $i$ and $j$.
Each segment consists of $j-i$ feature frames from a self-supervised learning (SSL) model (Fig.~\ref{fig:pipeline}b). 
After applying principal component analysis (PCA) to reduce the dimensionality, the SSL features are averaged across time to produce a single fixed-dimensional embedding $y_{i,j}$.
Although simple, this averaging approach for obtaining segmental representations has been shown to be highly effective~\cite{Levin2013, Jacobs2024, Herreilers2025}.
Fixed-dimensional representations are also useful for the next step, clustering.

\subsection{Clustering methods}
The final step in the term discovery pipeline is to cluster segments to form a lexicon (Fig.~\ref{fig:pipeline}c). 
This is the focus of this paper; we will use the same fixed-dimensional embeddings in all cases, and investigate the lexicons induced when altering the clustering method.\footnote{
Using fixed-dimensional representations allows different clustering approaches to be compared on equal footing. 
But two of the methods (graph clustering and agglomerative clustering) operate on pairwise similarities and are therefore compatible with variable-length representations, such as discrete unit sequences~\cite{vanNiekerk2024}. 
We nevertheless restrict all clustering methods to the same fixed-dimensional embeddings for a fair comparison. 
In preliminary experiments, we did also evaluate graph clustering using discrete SSL units, but observed no performance improvements over fixed-dimensional embeddings.
} 
Altogether, we compare five clustering methods: two bottom-up methods that operate on pairwise relationships between segment embeddings, and three centre-based methods that represent clusters through centroids or parametric components.
We describe the bottom-up methods~first.
\begin{figure}[!t]
\centering
\captionsetup{font=small}
\begin{tikzpicture}[node distance=4cm]
    \node (n1) [empty, anchor=west] at (0,0) {};
    \node[anchor=north, font=\small] at (-8.0, 2.5) {(a) Graph Construction};
    \begin{scope}
        \node[circle, draw, fill=gray!45, minimum size=0.2cm, inner sep=0pt] (g1) at (-10.0, 1.0) {};
        \node[circle, draw, fill=gray!45, minimum size=0.2cm, inner sep=0pt] (g2) at (-9.4, 1.5) {};
        \node[circle, draw, fill=gray!45, minimum size=0.2cm, inner sep=0pt] (g3) at (-8.9, 0.8) {};
        \node[circle, draw, fill=gray!45, minimum size=0.2cm, inner sep=0pt] (g4) at (-9.6, 0.4) {};
    
        \draw[thick] (g1) -- (g2);
        \draw[thick] (g2) -- (g3);
        \draw[thick] (g3) -- (g4);
        \draw[thick] (g1) -- (g4);
        \draw[thin, opacity=0.45] (g1) -- (g3);
        \draw[thin, opacity=0.45] (g2) -- (g4);

        \node[circle, draw, fill=gray!45, minimum size=0.2cm, inner sep=0pt] (g6) at (-8.5, 1.0) {};
        \node[circle, draw, fill=gray!45, minimum size=0.2cm, inner sep=0pt] (g7) at (-8, 1.5) {};
        \node[circle, draw, fill=gray!45, minimum size=0.2cm, inner sep=0pt] (g8) at (-7.7, 0.6) {};
        \node[circle, draw, fill=gray!45, minimum size=0.2cm, inner sep=0pt] (g9) at (-7.0, 1.7) {};
        \node[circle, draw, fill=gray!45, minimum size=0.2cm, inner sep=0pt] (g10) at (-6.9, 0.4) {};
        \node[circle, draw, fill=gray!45, minimum size=0.2cm, inner sep=0pt] (g11) at (-6.4, 1.0) {};

        \draw[thick] (g6) -- (g7);
        \draw[thick] (g7) -- (g8);
        \draw[thick] (g11) -- (g9);        
        \draw[thick] (g10) -- (g11);
        \draw[thin, opacity=0.45] (g10) -- (g9);
        \draw[thin, opacity=0.45] (g7) -- (g9);
        \draw[thin, opacity=0.45] (g8) -- (g10);

        \draw[dashed, <->] (g6) -- (g8);
        
        \node[anchor=south, font=\scriptsize] at (-8.3, -0.2) {Add edge if $s_{p,q} > \tau$};
    \end{scope}

    \node (n2) [empty, anchor=west] at (4.5, 0) {};
    \node[anchor=north, font=\small] at (-3.8, 2.5) {(b) Graph Partitioning};

    \begin{scope}[xshift=4.5cm]
        \node[circle, draw, fill=gray!45, minimum size=0.2cm, inner sep=0pt] (g1) at (-10.0, 1.0) {};
        \node[circle, draw, fill=gray!45, minimum size=0.2cm, inner sep=0pt] (g2) at (-9.4, 1.5) {};
        \node[circle, draw, fill=gray!45, minimum size=0.2cm, inner sep=0pt] (g3) at (-8.9, 0.8) {};
        \node[circle, draw, fill=gray!45, minimum size=0.2cm, inner sep=0pt] (g4) at (-9.6, 0.4) {};
    
        \draw[thick] (g1) -- (g2);
        \draw[thick] (g2) -- (g3);
        \draw[thick] (g3) -- (g4);
        \draw[thick] (g1) -- (g4);
        \draw[thin, opacity=0.45] (g1) -- (g3);
        \draw[thin, opacity=0.45] (g2) -- (g4);

        \node[circle, draw, fill=gray!45, minimum size=0.2cm, inner sep=0pt] (g6) at (-8.5, 1.0) {};
        \node[circle, draw, fill=gray!45, minimum size=0.2cm, inner sep=0pt] (g7) at (-8, 1.5) {};
        \node[circle, draw, fill=gray!45, minimum size=0.2cm, inner sep=0pt] (g8) at (-7.7, 0.6) {};
        \node[circle, draw, fill=gray!45, minimum size=0.2cm, inner sep=0pt] (g9) at (-7.0, 1.7) {};
        \node[circle, draw, fill=gray!45, minimum size=0.2cm, inner sep=0pt] (g10) at (-6.9, 0.4) {};
        \node[circle, draw, fill=gray!45, minimum size=0.2cm, inner sep=0pt] (g11) at (-6.4, 1.0) {};

        \draw[thick] (g6) -- (g7);
        \draw[thick] (g7) -- (g8);
        \draw[thick] (g11) -- (g9);        
        \draw[thick] (g10) -- (g11);
        \draw[thick] (g6) -- (g8);
        \draw[thin, opacity=0.45] (g10) -- (g9);
        \draw[dashed, opacity=0.45] (g7) -- (g9)
            node[midway, opacity=0.8, font=\scriptsize] {$\times$};
        \draw[dashed, opacity=0.45] (g8) -- (g10)
            node[midway, opacity=0.8, font=\scriptsize] {$\times$};
        
        \node[anchor=south, font=\scriptsize] at (-8.3, -0.2) {Repeat: if $\Delta Q>0$, reassign node};

        \fill[gray!30, opacity=0.25] (-9.48,0.95) ellipse [x radius=0.78, y radius=0.72];
        \fill[gray!30, opacity=0.25] (-8.05,1.05) ellipse [x radius=0.62, y radius=0.72];
        \fill[gray!30, opacity=0.25] (-6.78,1.03) ellipse [x radius=0.58, y radius=0.82];

    \end{scope}
        
\end{tikzpicture}
\caption{
    Graph construction using similarity threshold $\tau$ and partitioning via the Leiden algorithm, controlled by resolution parameter $\gamma$.
}
\vspace{-10pt}      
\label{fig:graph_steps}
\end{figure}
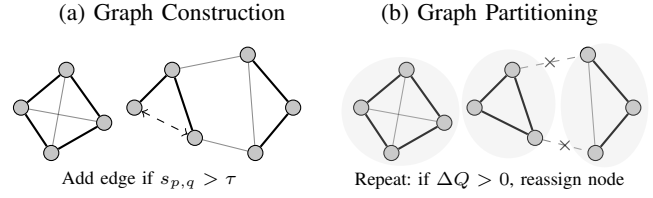

\textbf{Graph clustering:}
This approach proceeds in the two stages seen in Fig.~\ref{fig:graph_steps}.
First, a similarity graph $G = (V, E)$ is constructed (Fig.~\ref{fig:graph_steps}a). 
Each vertex $v \in V$ corresponds to a segment embedding $y_{i,j}$. 
For each pair of vertices $v_p$ and $v_q$, the corresponding cosine similarity $s_{p,q}$ is computed. 
If $s_{p,q} > \tau$, an edge $e \in E$ is added between $v_p$ and $v_q$ with weight $s_{p,q}$.
Higher values of this threshold parameter $\tau$ yield sparser graphs with smaller, more isolated communities, resulting in a flatter type-frequency~distribution.

The connected components of this thresholded graph form an initial partition (Fig.~\ref{fig:graph_steps}a).
However, this partition is coarse: segments may be grouped together through chains of pairwise similarities exceeding $\tau$, which can result in communities that contain pairs of segments that are not similar. 
In the second stage, we apply the Leiden algorithm~\cite{Traag2019} to refine this partitioning (Fig.~\ref{fig:graph_steps}b).
Leiden searches for a partition that locally maximises a quality function over the graph.
For this, we employ the constant Potts model (CPM)~\cite{Traag2011}:
\begin{equation}
Q = \sum_{p=1}^N \sum_{q=1}^N (s_{p,q} - \gamma)\cdot\delta(c_p, c_q)
\label{eq:cpm}
\end{equation}
\noindent where $N$ is the total number of vertices, $c_p$ and $c_q$ denote the community assignments for vertices $p$ and $q$, and $\delta(c_p,c_q)$ is an indicator equalling $1$ if the two vertices are assigned to the same community and $0$ otherwise.
The CPM objective rewards pairs of vertices assigned to the same community according to their edge weight $s_{p,q}$, while penalising such assignments by a fixed cost $\gamma$. 
The resolution parameter $\gamma$ therefore gives another control over cluster granularity: higher values of $\gamma$ favour smaller, more homogeneous clusters, while lower values allow larger, more inclusive clusters.
Importantly, $\gamma$ operates within the graph structure defined by $\tau$: it can further subdivide communities but cannot recover connections excluded by $\tau$.

Graph clustering, including Leiden with CPM, has been used previously for unsupervised term discovery~\cite{Slabbert2026}.
Here we extend~\cite{Slabbert2026} through a systematic multilingual, multi-granularity study of lexical distributional structure, including unsupervised boundaries.
Park and Glass~\cite{Park2008} used the older Newman algorithm~\cite{Newman2004} to partition the graph.
Later work by Lyzinski et al.~\cite{Lyzinski2015} applied the Louvain algorithm~\cite{Blondel2008}, a predecessor to Leiden.
They also optimised modularity rather than CPM.

We use Leiden with CPM because it is more robust to disconnected communities~\cite{Traag2019} and gives direct control over cluster granularity through $\gamma$.
This is important because we want the resulting clusters to follow a particular distribution.
In our own preliminary experiments, CPM also outperformed modularity when applied with the Leiden algorithm.

\textbf{Agglomerative clustering:}
Our second bottom-up method builds clusters hierarchically through successive merges.
At each step, the two clusters with the lowest average pairwise distance are joined, a criterion called average linkage~\cite{Johnson1967, Nielsen2016}.
The distance between two clusters $A$ and $B$ is defined as:
\begin{equation}
d(A, B) = \frac{1}{|A||B|} \sum_{a \in A} \sum_{b \in B} d(y_a, y_b)
\label{eq:avg_linkage}
\end{equation}
\noindent where $d(y_a,y_b)$ is the distance between embeddings $y_a$ and $y_b$.
Unlike Ward linkage~\cite{Murtagh2014}, which minimises within-cluster variance and therefore favours compact clusters around centroids, average linkage operates over average pairwise distances. 
This does, however, make it more computationally expensive, as it requires maintaining and updating pairwise distances between clusters at each merge step.
Rather than stopping merges at a distance threshold, we cut the hierarchy to obtain
a desired number of clusters $K$.
This is in contrast to graph clustering, where the parameters are adjusted to approximate the target~$K$.

\textbf{Centre-based methods:}
We compare against three centre-based methods that represent each cluster by a central point.
K-means partitions data into $K$ clusters by minimising within-cluster variance, implicitly favouring compact clusters around centroids.
BIRCH~\cite{Zhang1996} incrementally builds a tree of compact subclusters by absorbing each new point into the nearest subcluster within a fixed threshold; if no such subcluster exists, a new one is created. 
Since we provide a target $K$, BIRCH then applies agglomerative clustering 
to the subcluster centroids to obtain the final partition.
This makes BIRCH highly efficient, but also constrains it toward compact clusters, since the initial subcluster construction is based on a fixed threshold around local centroids.
The finite Bayesian Gaussian mixture model (FBGMM)~\cite{Rasmussen1999, Kamper2014} is a Bayesian variant of a Gaussian mixture model (GMM) that places a Dirichlet prior over mixture weights, allowing greater variation in cluster sizes than a standard GMM.
We include it as a representative probabilistic method and because of its prior use in unsupervised term discovery~\cite{Kamper2017es}. 
Although distributions over cluster means are modelled, the FBGMM is still fundamentally centre-based.

Together, these five methods allow us to test whether bottom-up approaches, which build clusters from pairwise similarities, are better suited than centre-based methods for recovering the long-tailed type-frequency structure of natural lexicons.

\section{Experimental Setup}
\label{sec:experimental_setup}
We evaluate the five clustering methods across three segmentation conditions (ground-truth words, ground-truth syllables, and unsupervised syllabic segments) and three languages (English, Afrikaans, and French).
This allows us to analyse sensitivity to segmentation choice and to assess whether the observed clustering behaviour is consistent across languages.

\textbf{Datasets:} 
For English, we use LibriSpeech dev-clean~\cite{Panayotov2015}, comprising 4.5 hours of active speech from 40 speakers. 
For Afrikaans, we use 2 hours of speech from five speakers drawn from FLEURS~\cite{Conneau2022}.
Afrikaans is a low-resource Germanic language (like English), spoken in southern Africa.
For French, we use a 4.2-hour subset (comparable to the English set) from Track~2 of the ZeroSpeech challenge~\cite{Dunbar2017}, covering 12 speakers. 
Although French is a high-resource language, we treat it here as unseen.
All three sets are relatively small, but these sizes are reasonable since no model training is taking place (an existing SSL model is used; see below).
Moreover, many of the clustering approaches are computationally expensive; our goal here is a fair comparison without limitations due to computational constraints (although efficiency is discussed~later).

\textbf{Representations:}
We follow the best setup from~\cite{Slabbert2026}.
Frame-level representations are extracted from the 21st layer of WavLM Large~\cite{Chen2022}. 
Extracted features are mean- and variance-normalised and reduced to 350 dimensions using PCA. 
Segment-level embeddings are then obtained by mean pooling over time. 
The resulting segment embeddings are mean-centred across the dataset and $l_2$-normalised. 
WavLM was trained on large amounts of English data (no Afrikaans or French).

\textbf{Segmentation conditions:}
Our bigger goal is to improve \textit{unsupervised} term discovery, but here our focus is specifically on the effects of the clustering step.
Given the limited performance of unsupervised word segmentation systems, we therefore start with ground-truth word boundaries.
Unsupervised systems for detecting syllabic boundaries have been argued to be much more reliable~\cite{Rsnen2015};
we therefore also consider both ground-truth and unsupervised syllabic boundaries.
Word and phone alignments for English and Afrikaans are obtained using a forced aligner~\cite{McAuliffe2017}, while the French alignments are provided with the data.
Syllable alignments for all three languages are derived from the phone alignments using rule-based methods.
Unsupervised syllable boundaries are inferred using the ZeroSyl method~\cite{Visser2026}, where boundaries are placed at prominent peaks in the smoothed $l_2$-norm signal of the WavLM features.

\textbf{Implementation:}
Graph clustering is implemented using \texttt{igraph}~\cite{Csardi2006}. 
K-means clustering is performed using FAISS~\cite{Douze2025}  with K-means++ initialisation~\cite{Arthur2007}. 
BIRCH and agglomerative clustering are implemented using \texttt{scikit-learn}~\cite{Pedregosa2011}. 
BIRCH is configured with default hyperparameters (branching factor $50$, threshold $0.5$), and cluster centroids are extracted directly from the leaf nodes.
For agglomerative clustering, we use Euclidean distance for $d(y_a,y_b)$ in~\eqref{eq:avg_linkage}. (This worked better than cosine.)
The FBGMM follows~\cite{Kamper2017es}, with a fixed spherical covariance $\sigma^2 \mathbf{I}$ for all clusters; we set $\sigma^2 = 0.1$ based on development experiments.

Automatically estimating the lexicon size $K$ is an open research question that we do not address here. We therefore set a target $K$ across all methods. 
For the ground-truth word or syllable segmentation conditions, $K$ is set to the true number of reference types.
For unsupervised syllabic units, we target $K=5\text{k}$ across languages, with $K=10\text{k}$ as a sensitivity check.
This is based on work in syllable-based spoken language modelling, where a fixed $K$ of roughly this order is used across datasets and languages~\cite{Peng2023, Cho2024, Cho2025}.
For methods that require a fixed number of clusters, the target $K$ is specified directly.
For graph clustering, the resolution parameter $\gamma$ is tuned so that the desired $K$ is roughly~reached.

Prior to partitioning using $\gamma$ (Fig.~\ref{fig:graph_steps}b), the initial graph needs to be constructed using the similarity threshold $\tau$ (Fig.~\ref{fig:graph_steps}a).
We set $\tau$ based on English development experiments with true word and syllable boundaries.
These experiments showed that lower thresholds perform better on word segments, while higher thresholds perform better on syllable segments.
We therefore use $\tau=0.3$ for words, which is the lowest value that remained computationally tractable, and $\tau=0.55$ for syllables,\footnote{An exception is made for the ground-truth French syllables, where we use $\tau=0.5$ to obtain a connected graph, as the higher threshold of $\tau=0.55$ resulted in a graph that was too sparse to partition into the target $K$.} which is the highest value that still yields a connected graph. 

\textbf{Evaluation:}
Apart from distribution plots, we need quantitative metrics to compare approaches.
The most widely used metric for lexicon evaluation is normalised edit distance (NED)~\cite{Ludusan2014}: Each discovered segment is mapped to the ground-truth phone sequence with which it overlaps most, and the phone edit distance is then calculated between all pairs in a cluster. The distances are then aggregated.
NED has recently been revised to account for an unfair bias that overweights larger clusters; we use this revised metric, the normalised edit similarity (NES)~\cite{Malan2026}.
Both NED and NES only evaluate consistency within a cluster, ignoring how true types are spread over clusters.
To measure this, we therefore also use the inverse NES (iNES), which measures how consistently instances of each ground-truth type are clustered into the same cluster~\cite{Malan2026}.\footnote{In~\cite{Malan2026}, these metrics are called weighted normalised edit similarity (WNES) and inverse WNES (iWNES) to explicitly distinguish them from the original NED.}
NES and iNES can be combined by taking the harmonic mean, giving $F_1$NES.
Higher is better for all metrics.
We also report bitrate~\cite{Dunbar2020}, the lower bound on the average number of bits required to encode the tokenised speech (lower is better).

\section{Results}
\subsection{Centre-based vs bottom-up clustering on English}

We begin with English experiments, comparing the five clustering methods across the three segmentation conditions, where the SSL features were trained on the target language.

Table~\ref{tab:english_comparison} shows that graph and agglomerative clustering substantially outperform K-means, BIRCH and the FBGMM, especially when considering the recall-based iNES and combined $F_1$NES metrics.
At the word level, centre-based methods achieve high NES of 88--89\% but low iNES of 26--34\%, indicating fragmentation of true types.
In comparison, agglomerative and graph clustering more than double iNES, yielding $F_1$NES scores of 66\% and 67\% compared to 40\% for K-means at the word level.
The same pattern holds for the ground-truth and unsupervised syllable-level conditions; although overall performance drops under unsupervised segmentation, the relative advantage of bottom-up methods persists.
Under all three segmentation conditions, agglomerative and graph clustering also achieve the lowest bitrate, indicating distributions that efficiently encode the data.

\begin{figure*}[t!]
    \centering
    \includegraphics[width=\linewidth]{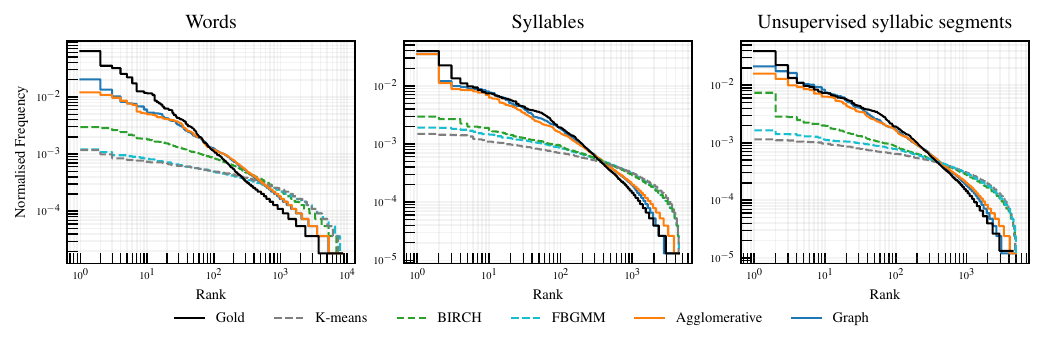}
    \captionsetup{font=small}
    \caption{
        Type-frequency distributions for induced lexicons on English across the three segmentation conditions. 
        Dashed lines correspond to centre-based methods while solid lines indicate bottom-up clustering techniques. 
        The black line shows the true word or syllable distribution.
        }
    \label{fig:zipf_plot_dev-clean-feature-sliced}
    \vspace{-10pt}      
\end{figure*}

\begin{table}[b!]
    \centering
    \captionsetup{font=small}
    \caption{
        Lexicon evaluation for English across three segmentation conditions.
        Bitrate is measured in bits/second.
    }
    \label{tab:english_comparison}
    \renewcommand{\arraystretch}{1.1}
    \begin{tabularx}{\linewidth}{@{}lCRRRR@{}}
    \toprule
    Method & $K$ & NES$\uparrow$ & iNES$\uparrow$ & $F_1$NES$\uparrow$ & Bitrate$\downarrow$ \\
    \midrule
    \multicolumn{6}{c}{Word-level ($K=8{,}372$)} \\
    \midrule
    K-means & 8{,}372 & 87.86 & 26.11 & 40.26 & 43.02 \\
    BIRCH & 8{,}372 & 88.12 & 33.67 & 48.72 & 41.79 \\
    FBGMM & 8{,}372 & 88.72 & 31.85 & 46.88 & 43.22\\
    Agglomerative & 8{,}372 & \textbf{91.31} & 52.71 & 66.83 & 38.50\\
    Graph ($\tau=0.3$) & 8{,}372 & 90.41 & \textbf{54.18} & \textbf{67.91} & \textbf{38.18} \\
    \midrule
    \multicolumn{6}{c}{Syllable-level ($K=4{,}473$)} \\
    \midrule
    K-means & 4{,}473 & \textbf{79.12} & 20.49 & 32.56 & 56.20 \\
    BIRCH & 4{,}473 & 78.82 & 23.39 & 36.07 & 55.57 \\
    FBGMM & 4{,}473 & 78.07 & 23.14 & 35.70 & 55.74\\
    Agglomerative & 4{,}473 & 77.08 & 41.50 & 53.96 & 49.52 \\
    Graph ($\tau=0.55$) & 4{,}473 & 74.78 & \textbf{44.44} & \textbf{55.75} & \textbf{34.30} \\
    \midrule
    \multicolumn{6}{c}{Unsupervised syllabic: ZeroSyl ($K=5\text{k}$)} \\
    \midrule
    K-means & 5{,}000 & \textbf{68.25} & 11.50 & 19.68 & 63.00 \\
    BIRCH & 5{,}000 & 67.55 & 13.53 & 22.55 & 61.97 \\
    FBGMM & 5{,}000 & 67.30 & 12.44 & 21.00 & 62.64\\
    Agglomerative & 5{,}000 & 64.87 & 24.06 & 35.11 & 55.34 \\
    Graph ($\tau=0.55$) & 4{,}999 & 62.51 & \textbf{26.29} & \textbf{37.01} & \textbf{53.22} \\
    \bottomrule
    \end{tabularx}
    \vspace{-10pt}   
\end{table}

Fig.~\ref{fig:zipf_plot_dev-clean-feature-sliced} shows that these improvements correspond to clear differences in induced lexicon structure.
K-means, BIRCH, and the FBGMM produce flat type-frequency distributions, consistent with fragmentation of frequent types (as per iNES). 
In contrast, graph and agglomerative clustering recover distributions that more closely follow the true distributions. 
Comparing the gold ground-truth distributions (solid black lines), we see that the long tail of true words is less pronounced at the syllable level (as expected). 
Regardless, the bottom-up graph and agglomerative methods are able to track the true distribution for both  word- and syllable-level lexicon discovery.

\begin{figure}[t!]
    \vspace{-10pt}      
    \centering
    \captionsetup{font=small}
        \includegraphics[width=\linewidth]{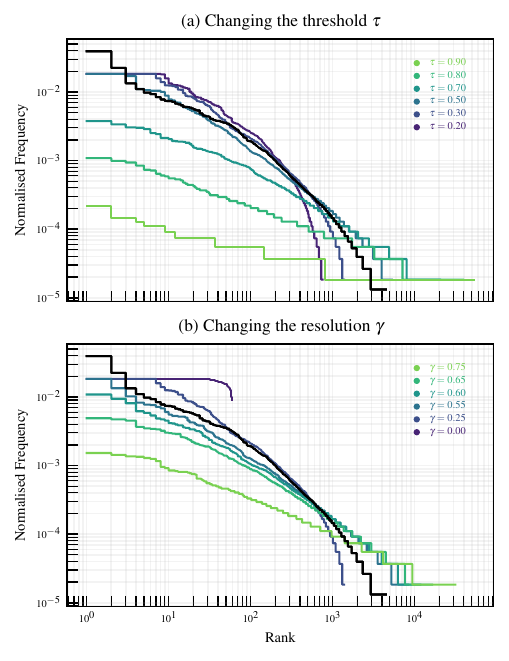}
        \caption{
            Effect of the (a) threshold $\tau$ and (b) resolution $\gamma$ graph clustering hyperparameters on the type-frequency distribution for true English word segments. 
            }
        \label{fig:zipf_plot_tau_gamma_dev}
        \vspace{-10pt}      
\end{figure}

\textbf{Graph vs\ agglomerative:}
Both bottom-up approaches provide a clear benefit, but there is no clear winner.
Although computational efficiency is not our core focus here, it is worth noting that agglomerative is much slower because of its repeated pairwise comparisons: on an 8-core 3.60 GHz CPU machine, agglomerative takes roughly 9 minutes compared to graph clustering's 3 minutes when clustering the true English word segments.
Agglomerative clustering also has only a single setting: determining when to cut the tree, while graph clustering gives more explicit control over the resulting distributions through its two~hyperparameters.

\begin{figure}[t!]
    \centering
    \captionsetup{font=small}
    \includegraphics[width=\linewidth]{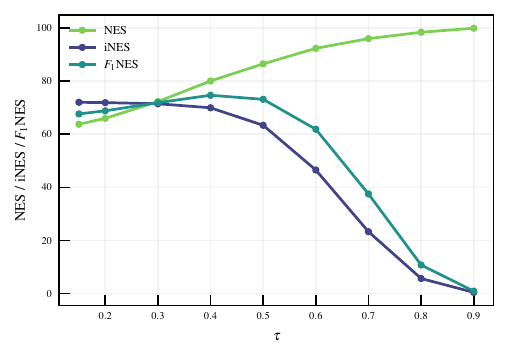}
    \caption{
        NES, iNES, and $F_1$NES at different thresholds $\tau$ for true English word segments.
    }
    \label{fig:tau_vs_nes}
    \vspace{-10pt}      
\end{figure}
   
\textbf{Distributions induced through graph clustering:}
Fig.~\ref{fig:zipf_plot_tau_gamma_dev} illustrates how the distributions change with the two graph clustering hyperparameters.
Higher $\tau$ yields sparser graphs with flatter, more uniform distributions better suited to syllables; lower $\tau$ produces denser graphs whose distributions more closely follow the true word distribution (consistent with optimal $\tau$ settings from development; see Sec.~\ref{sec:experimental_setup}).
The same trend holds for $\gamma$ (Fig.~\ref{fig:zipf_plot_tau_gamma_dev}b), but $\gamma$'s ability to approximate the true distribution is bounded by $\tau$: a sufficiently high $\tau$ could produce a graph too sparse for $\gamma$ to recover Zipfian structure.
Fig.~\ref{fig:tau_vs_nes} further shows how $\tau$ governs a trade-off between NES and iNES. 
At high~$\tau$, the graph fragments into many small communities, driving up NES while iNES and $F_1$NES collapse. 
At low~$\tau$, frequent types absorb neighbouring segments into large communities, improving iNES at the cost of NES. 
Together, $\tau$ and $\gamma$ therefore provide interpretable control over lexicon structure that goes beyond simply fixing $K$.

Taken together, the English results show that the clustering method affects both lexicon quality and the induced type-frequency distribution, and that bottom-up methods like graph clustering outperform centre-based methods like K-means.
The next section examines whether these advantages persist when applying English-trained SSL features on unseen languages.

\subsection{Centre-based vs bottom-up clustering with English-trained SSL features on Afrikaans and French}


\begin{table}[t!]
    \centering
    \captionsetup{font=small}
        \centering
        \caption{Lexicon evaluation for Afrikaans across three segmentation conditions.
        Bitrate is measured in bits/second.
        }
        \tablesep
        \label{tab:afrikaans_comparison}
        \renewcommand{\arraystretch}{1.1}
        \begin{tabularx}{\linewidth}{@{}lCRRRR@{}}
        \toprule
        Method & $K$ & NES$\uparrow$ & iNES$\uparrow$ & $F_1$NES$\uparrow$ & Bitrate$\downarrow$ \\
        \midrule
        \multicolumn{6}{c}{Word-level ($K=4{,}973$)} \\
        \midrule
        K-means	& 4{,}973 & 67.86 & 14.35 & 23.69 & 33.22 \\
        Agglomerative & 4{,}973 & \textbf{71.93} & \textbf{25.66} & \textbf{37.82} & 31.15 \\
        Graph ($\tau=0.3$) & 4{,}783 & 70.50 & 24.25 & 36.08 & \textbf{30.88}\\    
        \midrule
        \multicolumn{6}{c}{Syllable-level ($K=3{,}489$)} \\
        \midrule
        K-means & 3{,}489 & 66.11 & 13.64 & 22.61 & 48.44  \\
        Agglomerative & 3{,}489 & \textbf{68.37} & 28.04 & 39.77 & 44.18 \\
        Graph ($\tau=0.55$) & 3{,}496 & 59.26 & \textbf{41.26} & \textbf{48.65} & \textbf{37.08} \\
        \midrule
        \multicolumn{6}{c}{Unsupervised syllabic: ZeroSyl ($K=5\text{k}$)} \\
        \midrule
        K-means & 5{,}000 & \textbf{61.24} & 8.73 & 15.28 & 64.49 \\
        Agglomerative & 5{,}000 & 59.94 & 16.10 & 25.38 & 58.33 \\
        Graph ($\tau=0.55$) & 5{,}001 & 56.67 & \textbf{18.06} & \textbf{27.40} & \textbf{57.25} \\  
        \bottomrule
        \end{tabularx}
\end{table}

We repeat the experiments on Afrikaans and French data while still using the English-trained SSL model.
Here we compare the bottom-up approaches only to K-means, since BIRCH and FBGMM performed similarly to K-means on~English.

\begin{figure}[t!]
    \centering
    \vspace{-5pt}
    \captionsetup{font=small}
    \includegraphics[width=\linewidth]{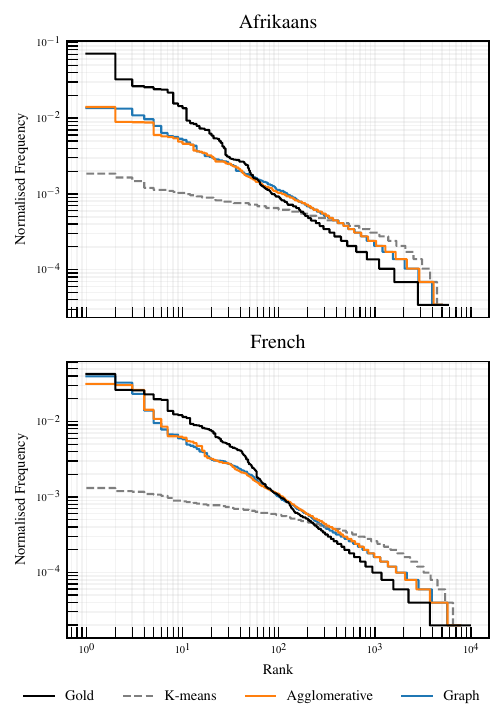}
    \caption{
        Type-frequency distributions for induced lexicons on Afrikaans and French data using ground-truth word segments.
    }
    \label{fig:zipf_word_level_two_languages}
    \vspace{-20pt}      
\end{figure}

Tables~\ref{tab:afrikaans_comparison} and~\ref{tab:french_comparison} show
that the same pattern of results observed for English holds for both Afrikaans and French, with graph and agglomerative clustering outperforming the standard centre-based K-means approach.
iNES scores increase substantially at the word level, from 15\% to 24--25\% on Afrikaans and from 10\% to 20--22\% on French.
At the syllable level, graph clustering 
has a clearer advantage over agglomerative clustering, particularly for Afrikaans, where it achieves the highest iNES and $F_1$NES at both syllable segmentations. 
For French, this advantage only holds for the ground-truth syllable segmentation.
Overall, performance degrades relative to English in both languages---more severely for French than for Afrikaans.
Fig.~\ref{fig:zipf_word_level_two_languages} confirms that graph and agglomerative clustering recover word-level type-frequency distributions closer to the reference distributions than K-means in both languages.
At $K=10\text{k}$, both bottom-up methods retain their $F_1$NES advantage over K-means across all three languages under unsupervised syllabic segmentation.


\begin{table}[t!]
        \centering
        \captionsetup{font=small}
        \caption{Lexicon evaluation for French across three segmentation conditions.
        Bitrate is measured in bits/second.}
        \tablesep
        \label{tab:french_comparison}
        \renewcommand{\arraystretch}{1.1}
        \begin{tabularx}{\linewidth}{@{}lCRRRR@{}}
        \toprule
        Method & $K$ & NES$\uparrow$ & iNES$\uparrow$ & $F_1$NES$\uparrow$ & Bitrate$\downarrow$ \\
        \midrule
        \multicolumn{6}{c}{Word-level ($K=7{,}776$)} \\
        \midrule
        K-means	& 7{,}776 & \textbf{65.66} & 9.91 & 17.22 & 40.97 \\
        Agglomerative & 7{,}776 & 65.32 & \textbf{22.17} & \textbf{33.10} & 36.36 \\
        Graph ($\tau=0.3$) & 7{,}776 & 62.92 & 20.42 & 30.83 & \textbf{35.73}\\    
        \midrule
        \multicolumn{6}{c}{Syllable-level ($K=2{,}522$)} \\
        \midrule
        K-means & 2{,}522 & \textbf{60.43} & 8.38 & 14.73 & 47.61\\
        Agglomerative & 2{,}522 & 52.24 & 20.99 & 29.95 & 38.79\\
        Graph ($\tau=0.5$) & 2{,}521 & 46.77 & \textbf{25.13} & \textbf{32.69} & \textbf{34.22}\\  
        \midrule
        \multicolumn{6}{c}{Unsupervised syllabic: ZeroSyl ($K=5\text{k}$)} \\
        \midrule
        K-means & 5{,}000 & \textbf{58.64} & 5.04 & 9.28 & 67.94 \\
        Agglomerative & 5{,}000 & 50.99 & 13.54 & 21.40 & 56.44 \\
        Graph ($\tau=0.55$) & 5{,}010 & 46.82 & \textbf{16.02} & \textbf{23.87} & \textbf{54.95} \\  
        \bottomrule
        \end{tabularx}
\end{table}

\section{Conclusion}

This paper examined the role of clustering in unsupervised term discovery.
We started by showing that the standard centre-based K-means approach has an inductive bias producing similarly sized clusters---a poor fit to the Zipfian distribution of real language.
Across three segmentation conditions and three languages, bottom-up methods (graph and agglomerative) consistently improved type-level recovery over centre-based algorithms (K-means, BIRCH, finite Bayesian GMM).
Graph clustering additionally provides interpretable control over the induced lexicon through two hyperparameters. 
Comparable performance from agglomerative clustering suggests that relaxing centre-based geometric assumptions, rather than the specific algorithm, is the key factor.
These findings establish clustering as a central modelling decision in unsupervised term discovery.
We found that graph clustering is computationally more efficient than agglomerative clustering, but both scale poorly. To be usable on realistically sized datasets, future work should focus on developing scalable bottom-up methods.

\bibliography{refs}

\end{document}